\documentclass[prx, reprint,longbibliography]{revtex4-1}

\draft
\usepackage[utf8]{inputenc}
\usepackage{graphicx}
\usepackage{times}
\usepackage{amsmath}
\usepackage{mathrsfs}
\usepackage{amssymb}
\usepackage{tensor}
\usepackage{braket}
\usepackage{mathptmx}
\usepackage{bm}
\usepackage{xcolor}
\usepackage{changes}

\usepackage{bbold}

\draft % marks overfull lines with a black rule on the right

\begin{document}

%\preprint{}

\title{Quantum-enhanced interferometry by entanglement-assisted rejection of environmental noise}

\begin{abstract}
Sensing and measurement tasks in severely adverse conditions such as loss, noise and dephasing can be improved by illumination with quantum states of light. Previous results have shown a modest reduction in the number of measurements necessary to achieve a given precision. Here, we compare three illumination strategies for estimating the relative phase in a noisy, lossy interferometer. When including a common phase fluctuation in the noise processes, we show that using an entangled probe achieves an advantage in parameter estimation precision that scales with the number of entangled modes. This work provides a theoretical foundation for the use of highly multimode entangled states of light for practical measurement tasks in experimentally challenging conditions. 
\end{abstract}

	\author{Alex O. C. Davis$^{1,2,3}$}
\email{aocd20@bath.ac.uk}

	\author{Giacomo Sorelli$^2$}

	\author{Val\'{e}rian Thiel$^{1,4}$}

	\author{Brian J. Smith$^{1,4}$}
		\affiliation{$^1$Clarendon Laboratory, University of Oxford, Parks Road, Oxford, OX1 3PU, UK}
\affiliation{$^2$Laboratoire Kastler Brossel, Sorbonne Universit\'{e}, CNRS, ENS-Universit\'{e} PSL, Coll\`{e}ge de France; 4 place Jussieu, F-75252 Paris, France}
\affiliation{$^3$Centre for Photonics and Photonic Materials, Department of Physics, University of Bath, Bath BA2 7AY, UK}
	\affiliation{$^4$Department of Physics and Oregon Center for Optical, Molecular, and Quantum Science, University of Oregon, Eugene, Oregon 97403, USA }
\date{\today}


\maketitle %\maketitle must follow title, authors, abstract and \pacs


% If in two-column mode, this environment will change to single-column format so that long equations can be displayed. 
%\begin{widetext}
%$$\mbox{put long equation here}$$
%\end{widetext}

% Figures should be put into the text as floats. 
% Use the graphics or graphicx packages (distributed with LaTeX2e).
% Put the label that you will use with \ref{} command in the braces of the \label{} command.
%
% \begin{figure}
% \end{figure}

% \begin{table}
% \end{tabular}
% \end{table}

%\begin{acknowledgments}
% Put your acknowledgments here.
%\end{acknowledgments}

%\bibliography{your-bib-file}

\section{Introduction}
In recent years, measurement schemes that use quantum states of light to overcome classical limitations on precision have attracted growing interest \cite{helstrom:76, giovannetti:11, pezze:14, paris:09}. Much of this research has focused on situations where precision is limited by \emph{intrinsic} noise of the probe input light such as shot noise, showing that nonclassical probe states achieve optical phase estimation with precision beyond the standard quantum limit \cite{sanders:95, motes:15, LIGO}.  However, the quantum states needed to realise these enhancements are often found not to be robust to extrinsic effects encountered in real-world applications such as loss and background noise. %Their potential is therefore restricted in scope, anticipating excellent control over experimental conditions or circumstances where the intensity of illumination must be strictly limited.

A separate approach to finding quantum enhancement in measurement problems instead considers situations where precision is limited by \emph{environmental} degradation such as loss, dephasing and background photons. Many real-world measurements take place under conditions where these effects are significant problems, and exploring quantum enhancements within this regime is a relatively new and evolving area of research. One result of this work is the quantum illumination (QI) protocol \cite{shapiro:20, sorelli:20}, which determines the existence of a reflecting target by probing through lossy channels in the presence of background noise. 
QI involves entangling the probe photon with an ancilla system retained by the observer. 
Unlike quantum phase estimation protocols intended to approach Heisenberg scaling, which are not robust to non-ideal conditions, QI yields a relative quantum advantage over classical illumination that actually increases with signal degradation, providing substantial enhancements even in the high-loss, high-noise limit.  

%While background noise shows no correlations with the ancilla, any remnant of the probe will still exhibit tell-tale correlations, even after signal degradation is so severe that the final state would fail any nonclassicality test. Hence QI exhibits quantum enhancements for a practically useful class of non-ideal target-detection problems. 
Lloyd's seminal work on QI \cite{lloyd:08} considers only such quantum radars with single-photon probes. 
It was subsequently shown that these can be outperformed by ``classical'' (coherent-state) probes \cite{shapiro:09}, which possess a well-defined phase that can be used to discern returned signal from the background noise. 
An advantage in the target detection error probability is recovered with full Gaussian quantum state illumination \cite{tan:08}, which has since been shown to be optimal for the target detection problem \cite{depalma:18,nair:20}. 
Somewhat disappointingly, this advantage represents only a constant factor of 4 in the error probability exponent, and does not scale with mode number.

\begin{figure*}
\includegraphics[width=0.8\linewidth]{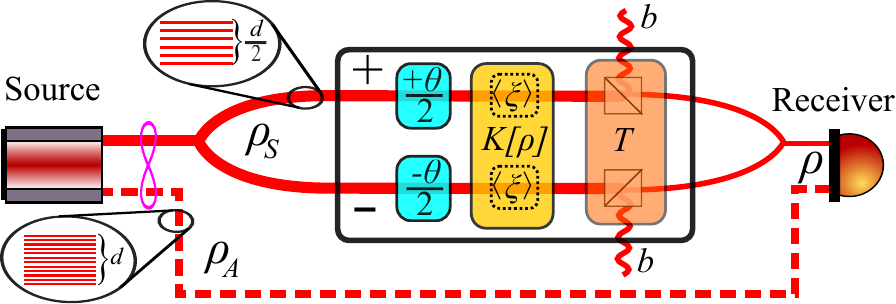}
\caption{Schematic diagram of measurement model. An optical probe $\rho_S$ is prepared across $d$ modes, half in the `+' subsystem, half in the `-' subsystem, and injected into the channel. Initially, the two sets of modes accumulate a relative phase difference of $\theta$, here modelled as $\pm\theta/2$ on each set. All modes then undergo a random, common phase fluctuation $\xi$, followed by coupling with transmissivity $T$ to noise modes with average photon number $b$ per mode. The output of the channel is then subject to optimal measurement at a receiver. In Case (2), the probe is also entangled with a $d$-mode ancilla system (bottom), which is available for the final measurement operation and whose presence enables discrimination against noise, providing a quantum enhancement.}
\label{channel}
\end{figure*}

Here we extend the concept of QI to the problem of phase estimation in noisy interferometry.
We consider the task of measuring the relative phase, $\theta$, between two sets of $d/2$ orthogonal lossy modes exposed to high background noise levels.
This measurement model is illustrated in Fig. \ref{channel}.
We compare three schemes with equal average photon flux corresponding to different probing strategies: (1) illumination of the channel with a single photon; (2) with a single photon maximally entangled with an observer-retained ancilla photon, and (3) the `semi-classical' case where illumination is by a coherent state with mean photon number 1. 
As in standard QI, we assume very substantial losses and background noise.
In a further departure from existing work on QI, we also allow the signal to be degraded by a random and uniform fluctuation in the optical path, giving rise to a common dephasing. 
By calculating the quantum Cramer-Rao bounds on measurement precision, we show that this common dephasing severly limits the effectiveness of classical illumination by eliminating the coherence between independent illumination trials, whilst quantum protocols retain their favourable scaling with mode number. We therefore find that the entangled scheme Case (2) achieves a large mode-dependent quantum enhancement over the semi-classical illumination of Case (3). 
%Here we consider a scenario where a target sample is illuminated by a channel comprising $d$ orthogonal probe modes, with a relative phase imparted on the modes. We compare the separable scenario, in which this channel contains a single photon in a superposition of these modes, with the entangled scenario, in which the probe is a single photon in an entangled state across the $d$ modes with a retained ancilla. The channel suffers loss and added noise, which we take to be equal for all modes.

%\begin{figure}[t]
%\centering
%\includegraphics[width=\linewidth]{falsepositive}
%\caption{ Conceptual schematic with $d$=2. A unitary quantum operation $U(\theta)$ is probed with an optical state $\Ket{+}$ (a) or $\Ket{\Psi -}$ (b). $U(\theta)$ distributes the signal into the output modes depending on $\theta$. A single output mode is monitored (blue). Background noise is randomly distributed across all output modes. Environmental noise in the detection mode give false positives (red crosses). In the entangled case (b), false positives in the detection mode occur less frequently relative to signal than the separable case (a).}
%\label{falspositive}
%\end{figure}

In the special case of single-photon probes, the origin of the quantum advantage in QI can be intuitively understood with the concept of false positive detection events. The target parameter determines the number of signal photons returned at the output of the measurement channel in the detection mode, with the accuracy of the measurement determined by the uncertainty in the number of counts in the detection mode that can be attributed to returned signal (true positives). Background environmental noise in the detection mode (false positives) cannot be distinguished from the signal, and so adds uncertainty to the parameter estimation.
In the unentangled case, the optimal scheme consists of sending a probe photon into the channel in one (known) mode of the $d$ modes available. These photons will then be coupled into the detection mode at the output of the channel, in proportions determined by the parameter of interest. 
% If $d=2$ such a basis is provided by the Bell states, defined by
%\begin{align}
%\Ket{\Phi\pm} &=\frac{1}{\sqrt{2}}\left( \Ket{0}_S\Ket{0}_A\pm\Ket{1}_S\Ket{1}_A\right) %\label{BellStates} \\
%\Ket{\Psi\pm}&=\frac{1}{\sqrt{2}}\left( \Ket{0}_S\Ket{1}_A\pm\Ket{1}_S\Ket{0}_A\right) \nonumber ,
%\end{align}
%where $\Ket{0}$ and $\Ket{1}$ denote the possible states of each of the pair of subsystems $A$ (ancilla) and $S$ (signal). 

By contrast, a maximally entangled photon pair exists in a Hilbert space with dimension $d^{2}$. In the QI scheme, a photon pair is prepared in one maximally entangled state across $d$ modes, and a joint measurement is performed on the ancilla photon and the photon returned from the channel. Since we assume ideal retention of the ancilla photon, the probability of a detection event is once again the probability of receiving a photon from the noisy channel. As there are no correlations between a background photon and the ancilla photon, noise-ancilla coincidence events will be randomly and uniformly distributed across the whole $d^{2}$-dimensional Hilbert space. This contrasts with the case of single-photon illumination, where the same number of background counts are distributed across only the $d$ modes of the measurement basis. The quantum enhancement thus derives from the fact that entanglement with an ancilla system distributes noise events more diffusely over a much larger space such that fewer occur in the detection modes. Importantly, this is a coherent effect, so classically correlated states cannot reproduce it. 

Coherent-state probes benefit from having contributions from multiple Fock states, which gives them a well-defined phase that can also be used to distinguish returned signal from background noise. We show that the presence of a common dephasing in the channel negates this advantage, creating a regime where entangled photon pairs show an advantage over classical illumination that scales with the number of modes of entanglement.

\section{Results} We consider illumination by a pure quantum state $\Ket{\psi}$ in the space $\mathcal{H}$. For separable illumination, we can write $\mathcal{H}=\mathcal{H}_S$, where $\mathcal{H}_S$ is the state space of the probe signal, whereas for ancilla-entangled probes, we instead have $\mathcal{H}=\mathcal{H}_S\otimes\mathcal{H}_A$, where $\mathcal{H}_A$ corresponds to the ancilla system and $\otimes$ indicates the tensor product. We can write the space of the signal as
\begin{equation}
\mathcal{H}_S=\left(\mathcal{H}^+_1\otimes...\otimes\mathcal{H}^+_{d/2}\right)\otimes \left(\mathcal{H}^-_1\otimes...\otimes\mathcal{H}^-_{d/2}\right),
\end{equation} 
where $\mathcal{H}^\pm_i$ indicates the Hilbert space of the $i$-th mode in the $\mathcal{H}^\pm$ subspace. Here $\mathcal{H}_S$ has been split into two subspaces $\mathcal{H}^+$ and $\mathcal{H}^-$ (with corresponding total photon number operators $\hat{n}^{+/-}$ and identities $\mathbb{1}^{+/-}$). The interaction to be measured is generated by a unitary operator 
\begin{equation}
\hat{U}_S(\theta)=\exp\{ i(\hat{n}^+\otimes\mathbb{1}^--\mathbb{1}^+\otimes\hat{n}^-)\theta/2\},
\end{equation} 
acting on $\ket{\psi}$, which simply adds a positive phase $\theta/2$ to the `$+$' modes and a negative phase $-\theta/2$ to the `$-$'  modes. The parameter $\theta$ is the target of the measurement. These two sets of modes may then be thought of as comprising two arms of an interferometer, with the target parameter being their relative phase.

We then include two sources of environmental degradation of the signal. Firstly, we assume strong coupling to a noisy background mode by a beam splitter transformation with transmissivity $T$. Secondly, we include a total global dephasing of the signal states $\hat{\rho}_S$ represented by a linear map,
\begin{equation}
\boldsymbol{K}[\hat{\rho}_S]=\frac{1}{2\pi} \int_{-\pi}^{\pi}e^{i\hat{n}\xi}\hat{\rho}_S e^{-i\hat{n}\xi}\text{d}\xi,
\label{dephasingop}
\end{equation}
where $\hat{n}\equiv\hat{n}^+\otimes\mathbb{1}^-+\mathbb{1}^+\otimes\hat{n}^-$ is the total photon number operator acting on $\mathcal{H}_S$. This source of noise is important in the quantum advantage, since it has no effect on (phase-insensitive) single-photon probes, but it randomises the global phase of coherent-state probes, leaving it unavailable as a distinguishing mark of the signal against the background noise and hence substantially degrading the precision with coherent-state probes. Since this random phase acts in common on all modes, it preserves the phase difference between the $+$ modes and the $-$ modes induced by $\hat{U}_S(\theta)$. Physically, our model corresponds to a lossy channel with both background noise and a common fluctuation in the optical path length of all modes, but with the difference in optical path between the $+/-$ modes remaining fixed (Fig. \ref{channel}). 

A real-world example of such a scenario is where $\theta$ is induced by the birefringence of a medium, $\pm$ gives the polarization state, and $i$ the index of another degree of freedom such as the spatial or temporal mode. This degree of freedom is not related to the target parameter, but is used as an ancilliary basis to enable discrimination against background noise. The global dephasing implies that the overall optical path fluctuates unpredictably on a scale much greater than a wavelength, but the relative phase between polarization modes ($\theta$) remains stable.

In Case (1), the probe $\Ket{\psi_1}$ is an eigenstate of $\hat{n}$ and is thus unaffected by the action of the dephasing superoperator $\boldsymbol{K}[\hat{\rho}_S]$ of Eq. \ref{dephasingop}. Hence the post-selected state of single photons detected at the channel output is given by the density matrix 
\begin{equation}
\hat{\rho}^1(\theta) = \frac{1-\eta}{d}\mathbb{1}_S + \eta \hat{U}_S(\theta)\Ket{\psi_1}\Bra{\psi_1}\hat{U}^{\dagger}(\theta).
\label{rho}
\end{equation}
The first term on the right of Eq. (\ref{rho}) is a maximally mixed density matrix corresponding to the background light, proportional to the identity operator $\mathbb{1}_S$ over $\mathcal{H}_S$.  $\eta$ is the fraction of photons at the receiver that originated from the initial illumination:
\begin{equation}
\eta=\frac{TN_\alpha}{TN_\alpha+N_\beta(1-T)},
\end{equation}
 where $N_\alpha$ is the average per-pulse photon number in the transmitted probe and $N_\beta$ is the average per-pulse photon number in the noise (summing over all modes). 
 We will assume that perfect single-photon probes are prepared with unit efficiency and hence that $N_\alpha=1$. Following Lloyd \cite{lloyd:08}, we will also use $N_\beta=bd$, with $b\ll1$ the average photon number in each noise mode prior to coupling with the signal mode. Hence we will later use the substitution
\begin{equation}
\eta=\frac{T}{T+bd(1-T)}.
\label{etaexpression}
\end{equation}

We next consider Case (2) where the input state $\Ket{\psi_2}$ consists of a single photon which is maximally entangled across all $d$ modes with a second $d$-mode ancilla photon that is ideally retained by the observer. In this case, photon pairs at the output of the channel have the state
\begin{equation}
\hat{\rho}^2(\theta) = \frac{1-\eta}{d^2}\mathbb{1}_S\otimes\mathbb{1}_A + \eta \hat{U}_{S\otimes A}(\theta)\Ket{\psi_2}\Bra{\psi_2}\hat{U}_{S\otimes A}^{\dagger}(\theta),
\label{rhoE}
\end{equation}
where $\mathbb{1}_A$ is the identity on $\mathcal{H}_A$ and $\hat{U}_{S\otimes A}(\theta)=\hat{U}_S(\theta)\otimes \mathbb{1}_A$.

Lastly we consider the semi-classical Case (3) where the probe $\Ket{\psi_3}$ consists of two coherent states on $\mathcal{H}^+$ and $\mathcal{H}^-$, each with mean photon number $|\alpha|^2=1/2$. We also model the background noise in each mode by a thermal state with mean photon number $b$. After the coupling to the background noise mode, the state may still be represented as a Gaussian state. However, this is no longer the case after phase randomization, and hence the final state can be written
\begin{equation}
\hat{\rho}^3(\theta)=\frac{1}{2\pi} \int_{-\pi}^{\pi}\hat{G}_{\delta,\sigma}^+(\xi+\theta/2)\otimes\hat{G}_{\delta,\sigma}^-(\xi-\theta/2)\text{d}\xi,
\label{semi-classicalrho}
\end{equation}
where $\hat{G}_{\delta,\sigma}^\pm(\xi)$ is a displaced thermal state on a mode in $\mathcal{H}^\pm$ with displacement $\delta=|\alpha|\sqrt{T}$, quadrature variance $\sigma^2=1/2+(1-T)b$ and phase $\xi$.

To compare the precision of the three schemes, we invoke the quantum Cramer-Rao bound \cite{helstrom:76, giovannetti:11, pezze:14, paris:09} (QCRB), which provides a measurement-independent lower bound on the variance $\sigma_\theta^2$ in the estimated value of the parameter $\theta$. It therefore yields the best possible precision attainable for a given probe state, optimised over all possible detection schemes. The QCRB, which is always saturable in the asymptotic limit $\nu \rightarrow \infty$, where $\nu$ is the number of experimental trials, is expressed as
\begin{equation}
\sigma_\theta^2 \geq \frac{1}{\nu I_Q}.
\end{equation}
Here $I_Q$ is the quantum Fisher information \cite{braunstein:94} per experimental trial, which expresses the maximum possible information about the target parameter that is attainable per measurement trial, for a given probe state. Here we derive the form of the quantum Fisher information for the three cases.

To calculate the QCRB for cases (1) and (2), where we have a discrete-variable (single-photon) probe, we introduce the basis of single-photon states $\Ket{\phi_j^\pm} \in \mathcal{H}_S$ (with $1\leq j \leq d/2$) which have a single photon on the $j$-th mode in the $\mathcal{H}^\pm$ subspace and zero photons on all other modes. The action of $\hat{U}_S(\theta)$ on these states $\Ket{\phi_j^\pm}$ is then
\begin{equation}
\hat{U}_S(\theta)\Ket{\phi_j^\pm}=\begin{cases}
 e^{i\theta/2}\Ket{\phi_j^+} \approx (1+i\theta/2)\Ket{\phi_{j}^+}\label{Udef}\\
 e^{-i\theta/2}\Ket{\phi_j^-} \approx (1-i\theta/2)\Ket{\phi_{j}^-}. 
\end{cases}
\end{equation}

The quantum Fisher information per detected photon is given by $\tilde{I}_Q=\mbox{Tr}[\hat{\rho}' R^{-1}_\rho(\hat{\rho}')]$ \cite{braunstein:94}, where $\hat{\rho}'=\partial\hat{\rho}/\partial\theta$ is the derivative of $\hat{\rho}$ with respect to $\theta$. Here $R^{-1}_\rho(A)$ is a superoperator acting on an operator $A$ such that 

%\begin{equation}
%[R^{-1}_\rho(A)]^{\pm,\pm}_{ij}=\frac{A_{ij}^{\pm,\pm}}{\beta_i+\beta_j}
%\end{equation}
%where $A_{ij}^{\pm,\pm}=\Bra{\phi^\pm_i}A\Ket{\phi^{\pm}_j}$ \replaced[id = GS]{with}{and close to $\theta=0$},
%\begin{equation}
%\hat{\rho} = \sum_{i=1}^{d}\beta_i %\Ket{\phi^\pm_i}\Bra{\phi^{\pm}_i}.
%\label{betas}
%\end{equation}

\begin{equation}
[R^{-1}_\rho(A)]_{ij}=\frac{A_{ij}}{\beta_i+\beta_j}
\end{equation}
with $A_{ij}=\Bra{e_i}A\Ket{e_j}$ the matrix elements of $A$ in the basis where the density matrix $\hat{\rho}(\theta = 0)$ is diagonal
\begin{equation}
\hat{\rho}(\theta = 0) = \sum_i\beta_i \Ket{e_i}\Bra{e_i}.
\label{betas}
\end{equation}

\subsection{ Case (1): Separable single-photon probes}

 Firstly, we consider the separable case, where the target is illuminated by an optimal probe state $\Ket{\psi}=\Ket{\uparrow}$, where $\Ket{\uparrow}=\frac{1}{\sqrt{2}}(\Ket{\phi^+_i}+\Ket{\phi^-_j})$ and $\Ket{\downarrow}=\frac{1}{\sqrt{2}}(\Ket{\phi^+_i}-\Ket{\phi^-_j})$ for some $i,j$.
 
 By considering the action of $\hat{U}_S(\theta)$ on these states, we can show \textbf{(Appendix 1)} that the quantum Fisher information per output photon is
\begin{equation}
\tilde{I}_Q^1=\mbox{Tr}(\hat{\rho}'R^{-1}_{\hat{\rho}^1}(\hat{\rho}'))=\frac{\eta^2 d}{4(1-\eta)+2\eta d}.
\label{peroutputphoton}
\end{equation}

Substituting in Eq. \ref{etaexpression} for $\eta$ and normalising by the probability of detecting a photon,
\begin{equation}
    p=T+bd(1-T)~,
\end{equation}
we obtain the quantum Fisher information of each illumination trial:

\begin{equation}\boxed{
I_Q^1=\frac{T^2}{2(T-(1-T)b)}
}
\label{fisher}
\end{equation}
This expression has no dependence on $d$, since the optimal measurement consists solely of counting photons in the states $\Ket{\uparrow}$ and $\Ket{\downarrow}$, with the other $d-2$ modes being ignored.

\subsection{ Case (2): Single-photon probes entangled with a retained ancilla}

We then imagine that the probe consists of a single photon entangled with an observer-retained ancilla system over $d$ modes. Here, the probe $\Ket{\psi_2}$ is instead drawn from a basis of maximally entangled states $\Ket{k,m}$

\begin{align}
&\Ket{k,m}=\frac{1}{\sqrt{d}}\sum^d_{j} e^{2i\pi jk /d}\Ket{\phi_j}_S\Ket{\chi_{j+m}}_A,
\end{align}
where $\Ket{\chi_i}_A$ for $1\leq i \leq d$ forms a basis over the ancilla system and
$\Ket{\phi_j}_S= \Ket{\phi^-_{j/2}}$ for $j\text{ even}$ and $\Ket{\phi^+_{(j+1)/2}}$ for $j\text{ odd}$.

The state of returned single photons can thus be written
\begin{align}
\hat{\rho}^E =\frac{(1-\eta)}{d^2}\mathbb{1}_S\otimes\mathbb{1}_A + \eta \hat{U}_{S\otimes A}(\theta)\Ket{k,m}\Bra{k,m}\hat{U}_{\otimes A}^{\dagger}(\theta) \nonumber
\end{align}

%where $\mbox{Tr}_S$ refers to the partial trace over the signal subsystem and the second equality follows from the fact that, since $\Ket{k,m}$ is maximally entangled, $\mbox{Tr}_S(\Ket{k,m}\Bra{k,m})=\mathbb{1}_A/d$, where $\mathbb{1}_A$ is the identity in the ancilla subsystem.

Again, by considering the action of $\hat{U}_{S\otimes A}$ on $\Ket{k,m}$ \textbf{(Appendix 2)} we find that the quantum Fisher information per detected pair in the entangled case, $I_Q^2$, is given by

\begin{equation} 
\tilde{I}_Q^2=\frac{d^2\eta^2}{4(1-\eta)+2\eta d^2}.
\label{pairfisher}
\end{equation}

Once more, we substitute in Eq. \ref{etaexpression} and normalise to the count probability to obtain the quantum Fisher information per trial:
\begin{equation}\boxed{
I_Q^2=\frac{T^2}{2(T-(1-T)b/d)}
}
\end{equation}
In contrast with Eq. \ref{fisher}, this expression contains a dependence on the number of modes of entanglement $d$, corresponding to the ancilla-assisted rejection of environmental noise counts from the detection modes.

\subsection{ Case (3): Coherent state probes}
Lastly, we consider the semi-classical comparison, where the interferometer is illuminated by a coherent state of average photon number 1 in a mode denoted $\uparrow$ (or $\downarrow$) that is an equal-weighted superposition of a mode in `+' and one in `-'. In this case, the state at the output is given by Eq. (\ref{semi-classicalrho}). We use a different approach to calculating the QCRB, using the expression for the quantum Fisher information in terms of the Bures distance \cite{uhlmann:92}:

\begin{equation}
I_Q(\theta=0)=4\left(\left.\frac{\partial d_\text{Bures}(\hat{\rho}(0),\hat{\rho}(\epsilon))}{\partial\epsilon}\right|_{\epsilon=0}\right)^2.
\end{equation}
The Bures distance constitutes a measure of distinguishability between two states and can be defined in terms of the Uhlmann fidelity \cite{uhlmann:76} $F(\hat{\rho}_1,\hat{\rho}_2)=\left(\text{Tr}\left[\sqrt{\sqrt{\hat{\rho}_1}\hat{\rho}_2\sqrt{\hat{\rho}_1}}\right]\right)^2$:

\begin{equation}
d_\text{Bures}(\hat{\rho}_1,\hat{\rho}_2)^2=2(1-F(\hat{\rho}_1,\hat{\rho}_2))
\end{equation}

Hence the quantum Fisher information can be represented in terms of a limit:

\begin{equation}
I_Q^3=8\lim_{\varepsilon \to 0} \frac{1-\text{Tr}\left[\sqrt{\sqrt{\hat{\rho}(0)}\hat{\rho}(\varepsilon)\sqrt{\hat{\rho}(0)}}\right]}{\varepsilon^2}
\label{iqlimit}
\end{equation}

\begin{figure}
\includegraphics[width=\linewidth]{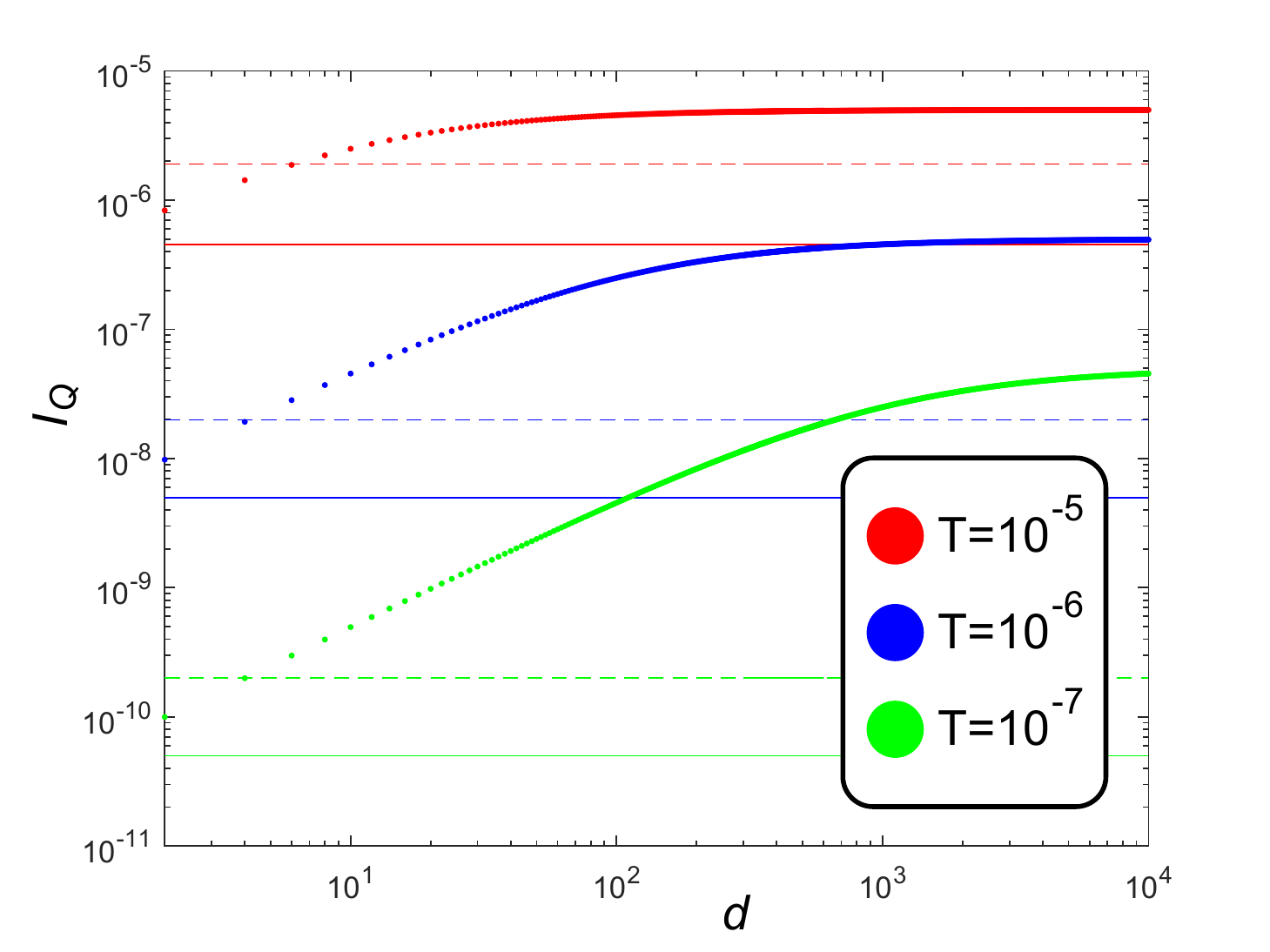}
\caption{Logarithmic plot showing the dependence of the quantum Fisher information on entanglement dimensionality $d$ for selected values of transmissivity $T$ and mean background photon number $b=10^{-4}$. The quantum Fisher information for entangled photon illumination $I_Q^2$ is plotted pointwise at even values of $d$. Also shown for the same values of $T$ and $b$ are the quantum Fisher information values for single-photon illumination ($I_Q^1$, solid horizontal lines) and coherent-state illumination ($I_Q^3$, dashed horizontal lines), which have no dependence on $d$. Note the asymptotic value of the quantum advantage obtained at $d\rightarrow \infty$ is proportionally greater for poorer transmission.}
\label{fig:enhancement}
\end{figure}

We derive an explicit expression for this limit \textbf{(see Appendix 3)}, which we evaluate numerically and compare with the other schemes in Fig. \ref{fig:enhancement}. Whilst it is generally greater than $I_Q^1$, the Fisher information for unentangled single-photon probes, it is also independent of $d$, since the optimal measurement again consists simply of monitoring the $\uparrow$ and $\downarrow$ modes. As such, for $b=10^{-4}$ and selected values of $T$ (see Fig. \ref{fig:enhancement}), the entangled scheme Case (2) outperforms the coherent state probe Case (3) for $d>4$. This advantage continues to grow as $d$ increases, eventually levelling off at $I_Q^2/I_Q^3 \sim b/T$ as $d\rightarrow \infty$. Physically, this corresponds to the dilution of the background counts from the detection mode of the entangled measurement to the point of insignificance, such that the Case (2) measurement is limited by shot noise in the returned signal, whilst the Cases (1) and (3) measurements remain limited by environmental noise. The maximum achievable quantum advantage is therefore greatest when both background noise and loss are severe.

\section{Conclusions} In summary, we have shown theoretically how the advantage gained by entanglement-enhanced rejection of extrinsic noise applies to estimating a continuous parameter, deriving a result showing that a large enhancement is accessible in a lossy and noisy regime. Interestingly, this speedup increases at higher values of loss, in stark contrast with protocols designed to gain a quantum advantage in the presence of intrinsic noise (e.g. shot noise), in which even modest losses can rapidly destroy quantum enhancements. Due to its innate resilience to adverse environmental conditions, this protocol holds the potential for an early real-world application of entanglement-enhanced technology to practical problems in communications and metrology.
Unlike previous results in QI, this advantage scales favourably with mode number due to the inclusion of common dephasing in the noise model, which undermines the use of phase as a signal/noise discriminator for classical probes. This suggests that the use of highly entangled optical probes could be used for a range of sensing or measurement tasks that are classically impractical due to noise including in LIDAR, atmospheric studies, or the characterization of optically thick media. The most important impediment to near-term application of this advantage is the lack of a suitable receiver for the entangled-basis measurements necessary to realise the quantum advantage. Issues with sources of entangled photons, such as their probabilistic nature or low brightness and purity, are another consideration. Future work in this framework could focus on either optimising the advantage by considering other types of probe state, such as two-mode squeezed states or entangled pairs of coherent states, or the effects of only partial common dephasing. \linebreak

\section{Acknowledgements} This project has received funding from the United Kingdom Defense Science and Technology
Laboratory (DSTL) under contract No. DSTLX-100092545. This work was partially funded by French ANR under COSMIC project (ANR-19-ASTR0020-01).

\bibliography{Bibliography}

\begin{thebibliography}{17}%
\makeatletter
\providecommand \@ifxundefined [1]{%
 \@ifx{#1\undefined}
}%
\providecommand \@ifnum [1]{%
 \ifnum #1\expandafter \@firstoftwo
 \else \expandafter \@secondoftwo
 \fi
}%
\providecommand \@ifx [1]{%
 \ifx #1\expandafter \@firstoftwo
 \else \expandafter \@secondoftwo
 \fi
}%
\providecommand \natexlab [1]{#1}%
\providecommand \enquote  [1]{``#1''}%
\providecommand \bibnamefont  [1]{#1}%
\providecommand \bibfnamefont [1]{#1}%
\providecommand \citenamefont [1]{#1}%
\providecommand \href@noop [0]{\@secondoftwo}%
\providecommand \href [0]{\begingroup \@sanitize@url \@href}%
\providecommand \@href[1]{\@@startlink{#1}\@@href}%
\providecommand \@@href[1]{\endgroup#1\@@endlink}%
\providecommand \@sanitize@url [0]{\catcode `\\12\catcode `\$12\catcode
  `\&12\catcode `\#12\catcode `\^12\catcode `\_12\catcode `\%12\relax}%
\providecommand \@@startlink[1]{}%
\providecommand \@@endlink[0]{}%
\providecommand \url  [0]{\begingroup\@sanitize@url \@url }%
\providecommand \@url [1]{\endgroup\@href {#1}{\urlprefix }}%
\providecommand \urlprefix  [0]{URL }%
\providecommand \Eprint [0]{\href }%
\providecommand \doibase [0]{http://dx.doi.org/}%
\providecommand \selectlanguage [0]{\@gobble}%
\providecommand \bibinfo  [0]{\@secondoftwo}%
\providecommand \bibfield  [0]{\@secondoftwo}%
\providecommand \translation [1]{[#1]}%
\providecommand \BibitemOpen [0]{}%
\providecommand \bibitemStop [0]{}%
\providecommand \bibitemNoStop [0]{.\EOS\space}%
\providecommand \EOS [0]{\spacefactor3000\relax}%
\providecommand \BibitemShut  [1]{\csname bibitem#1\endcsname}%
\let\auto@bib@innerbib\@empty
%</preamble>
\bibitem [{\citenamefont {Helstrom}\ and\ \citenamefont
  {Helstrom}(1976)}]{helstrom:76}%
  \BibitemOpen
  \bibfield  {author} {\bibinfo {author} {\bibfnamefont {Carl~W}\ \bibnamefont
  {Helstrom}}\ and\ \bibinfo {author} {\bibfnamefont {Carl~W}\ \bibnamefont
  {Helstrom}},\ }\href@noop {} {\emph {\bibinfo {title} {Quantum detection and
  estimation theory}}},\ Vol.~\bibinfo {volume} {3}\ (\bibinfo  {publisher}
  {Academic press New York},\ \bibinfo {year} {1976})\BibitemShut {NoStop}%
\bibitem [{\citenamefont {Giovannetti}\ \emph {et~al.}(2011)\citenamefont
  {Giovannetti}, \citenamefont {Lloyd},\ and\ \citenamefont
  {Maccone}}]{giovannetti:11}%
  \BibitemOpen
  \bibfield  {author} {\bibinfo {author} {\bibfnamefont {Vittorio}\
  \bibnamefont {Giovannetti}}, \bibinfo {author} {\bibfnamefont {Seth}\
  \bibnamefont {Lloyd}}, \ and\ \bibinfo {author} {\bibfnamefont {Lorenzo}\
  \bibnamefont {Maccone}},\ }\bibfield  {title} {\enquote {\bibinfo {title}
  {Advances in quantum metrology},}\ }\href@noop {} {\bibfield  {journal}
  {\bibinfo  {journal} {Nature Photonics}\ }\textbf {\bibinfo {volume} {5}},\
  \bibinfo {pages} {222} (\bibinfo {year} {2011})}\BibitemShut {NoStop}%
\bibitem [{\citenamefont {Pezzè}\ and\ \citenamefont
  {Smerzi}(2014)}]{pezze:14}%
  \BibitemOpen
  \bibfield  {author} {\bibinfo {author} {\bibfnamefont {Luca}\ \bibnamefont
  {Pezzè}}\ and\ \bibinfo {author} {\bibfnamefont {Augusto}\ \bibnamefont
  {Smerzi}},\ }\bibfield  {title} {\enquote {\bibinfo {title} {Quantum theory
  of phase estimation},}\ }in\ \href {\doibase 10.3254/978-1-61499-448-0-691}
  {\emph {\bibinfo {booktitle} {Proceedings of the International School of
  Physics "Enrico Fermi"}}},\ \bibinfo {series and number} {\bibinfo {number}
  {Course 188, Varenna}},\ \bibinfo {editor} {edited by\ \bibinfo {editor}
  {\bibfnamefont {G.~M.}\ \bibnamefont {Tino}}\ and\ \bibinfo {editor}
  {\bibfnamefont {M.~A.}\ \bibnamefont {Kasevich}}}\ (\bibinfo  {publisher}
  {IOS Press, Amsterdam},\ \bibinfo {year} {2014})\ pp.\ \bibinfo {pages} {691
  -- 741}\BibitemShut {NoStop}%
\bibitem [{\citenamefont {Paris}(2009)}]{paris:09}%
  \BibitemOpen
  \bibfield  {author} {\bibinfo {author} {\bibfnamefont {Matteo G.~A.}\
  \bibnamefont {Paris}},\ }\bibfield  {title} {\enquote {\bibinfo {title}
  {Quantum estimation for quantum technology},}\ }\href {\doibase
  10.1142/S0219749909004839} {\bibfield  {journal} {\bibinfo  {journal}
  {International Journal of Quantum Information}\ }\textbf {\bibinfo {volume}
  {07}},\ \bibinfo {pages} {125--137} (\bibinfo {year} {2009})}\BibitemShut
  {NoStop}%
\bibitem [{\citenamefont {Sanders}\ and\ \citenamefont
  {Milburn}(1995)}]{sanders:95}%
  \BibitemOpen
  \bibfield  {author} {\bibinfo {author} {\bibfnamefont {B.~C.}\ \bibnamefont
  {Sanders}}\ and\ \bibinfo {author} {\bibfnamefont {G.~J.}\ \bibnamefont
  {Milburn}},\ }\bibfield  {title} {\enquote {\bibinfo {title} {Optimal quantum
  measurements for phase estimation},}\ }\href {\doibase
  10.1103/PhysRevLett.75.2944} {\bibfield  {journal} {\bibinfo  {journal}
  {Phys. Rev. Lett.}\ }\textbf {\bibinfo {volume} {75}},\ \bibinfo {pages}
  {2944--2947} (\bibinfo {year} {1995})}\BibitemShut {NoStop}%
\bibitem [{\citenamefont {Motes}\ \emph {et~al.}(2015)\citenamefont {Motes},
  \citenamefont {Olson}, \citenamefont {Rabeaux}, \citenamefont {Dowling},
  \citenamefont {Olson},\ and\ \citenamefont {Rohde}}]{motes:15}%
  \BibitemOpen
  \bibfield  {author} {\bibinfo {author} {\bibfnamefont {Keith~R}\ \bibnamefont
  {Motes}}, \bibinfo {author} {\bibfnamefont {Jonathan~P}\ \bibnamefont
  {Olson}}, \bibinfo {author} {\bibfnamefont {Evan~J}\ \bibnamefont {Rabeaux}},
  \bibinfo {author} {\bibfnamefont {Jonathan~P}\ \bibnamefont {Dowling}},
  \bibinfo {author} {\bibfnamefont {S~Jay}\ \bibnamefont {Olson}}, \ and\
  \bibinfo {author} {\bibfnamefont {Peter~P}\ \bibnamefont {Rohde}},\
  }\bibfield  {title} {\enquote {\bibinfo {title} {Linear optical quantum
  metrology with single photons: exploiting spontaneously generated
  entanglement to beat the shot-noise limit},}\ }\href {\doibase
  10.1103/physrevlett.114.170802} {\bibfield  {journal} {\bibinfo  {journal}
  {Physical review letters}\ }\textbf {\bibinfo {volume} {114}},\ \bibinfo
  {pages} {170802} (\bibinfo {year} {2015})}\BibitemShut {NoStop}%
\bibitem [{\citenamefont {{LIGO collaboration}}(2013)}]{LIGO}%
  \BibitemOpen
  \bibfield  {author} {\bibinfo {author} {\bibnamefont {{LIGO
  collaboration}}},\ }\bibfield  {title} {\enquote {\bibinfo {title} {Enhanced
  sensitivity of the {LIGO} gravitational wave detector by using squeezed
  states of light},}\ }\href@noop {} {\bibfield  {journal} {\bibinfo  {journal}
  {Nature Photon.}\ }\textbf {\bibinfo {volume} {7}},\ \bibinfo {pages}
  {613--619} (\bibinfo {year} {2013})}\BibitemShut {NoStop}%
\bibitem [{\citenamefont {{Shapiro}}(2020)}]{shapiro:20}%
  \BibitemOpen
  \bibfield  {author} {\bibinfo {author} {\bibfnamefont {J.~H.}\ \bibnamefont
  {{Shapiro}}},\ }\bibfield  {title} {\enquote {\bibinfo {title} {The quantum
  illumination story},}\ }\href {\doibase 10.1109/MAES.2019.2957870} {\bibfield
   {journal} {\bibinfo  {journal} {IEEE Aerospace and Electronic Systems
  Magazine}\ }\textbf {\bibinfo {volume} {35}},\ \bibinfo {pages} {8--20}
  (\bibinfo {year} {2020})}\BibitemShut {NoStop}%
\bibitem [{\citenamefont {Sorelli}\ \emph {et~al.}(2020)\citenamefont
  {Sorelli}, \citenamefont {Treps}, \citenamefont {Grosshans},\ and\
  \citenamefont {Boust}}]{sorelli:20}%
  \BibitemOpen
  \bibfield  {author} {\bibinfo {author} {\bibfnamefont {Giacomo}\ \bibnamefont
  {Sorelli}}, \bibinfo {author} {\bibfnamefont {Nicolas}\ \bibnamefont
  {Treps}}, \bibinfo {author} {\bibfnamefont {Fr{\'e}d{\'e}ric}\ \bibnamefont
  {Grosshans}}, \ and\ \bibinfo {author} {\bibfnamefont {Fabrice}\ \bibnamefont
  {Boust}},\ }\bibfield  {title} {\enquote {\bibinfo {title} {Detecting a
  target with quantum entanglement},}\ }\href@noop {} {\bibfield  {journal}
  {\bibinfo  {journal} {arXiv preprint arXiv:2005.07116}\ } (\bibinfo {year}
  {2020})}\BibitemShut {NoStop}%
\bibitem [{\citenamefont {Lloyd}(2008)}]{lloyd:08}%
  \BibitemOpen
  \bibfield  {author} {\bibinfo {author} {\bibfnamefont {Seth}\ \bibnamefont
  {Lloyd}},\ }\bibfield  {title} {\enquote {\bibinfo {title} {Enhanced
  sensitivity of photodetection via quantum illumination},}\ }\href@noop {}
  {\bibfield  {journal} {\bibinfo  {journal} {Science}\ }\textbf {\bibinfo
  {volume} {321}},\ \bibinfo {pages} {1463--1465} (\bibinfo {year}
  {2008})}\BibitemShut {NoStop}%
\bibitem [{\citenamefont {Shapiro}\ and\ \citenamefont
  {Lloyd}(2009)}]{shapiro:09}%
  \BibitemOpen
  \bibfield  {author} {\bibinfo {author} {\bibfnamefont {Jeffrey~H}\
  \bibnamefont {Shapiro}}\ and\ \bibinfo {author} {\bibfnamefont {Seth}\
  \bibnamefont {Lloyd}},\ }\bibfield  {title} {\enquote {\bibinfo {title}
  {Quantum illumination versus coherent-state target detection},}\ }\href@noop
  {} {\bibfield  {journal} {\bibinfo  {journal} {New Journal of Physics}\
  }\textbf {\bibinfo {volume} {11}},\ \bibinfo {pages} {063045} (\bibinfo
  {year} {2009})}\BibitemShut {NoStop}%
\bibitem [{\citenamefont {Tan}\ \emph {et~al.}(2008)\citenamefont {Tan},
  \citenamefont {Erkmen}, \citenamefont {Giovannetti}, \citenamefont {Guha},
  \citenamefont {Lloyd}, \citenamefont {Maccone}, \citenamefont {Pirandola},\
  and\ \citenamefont {Shapiro}}]{tan:08}%
  \BibitemOpen
  \bibfield  {author} {\bibinfo {author} {\bibfnamefont {Si-Hui}\ \bibnamefont
  {Tan}}, \bibinfo {author} {\bibfnamefont {Baris~I}\ \bibnamefont {Erkmen}},
  \bibinfo {author} {\bibfnamefont {Vittorio}\ \bibnamefont {Giovannetti}},
  \bibinfo {author} {\bibfnamefont {Saikat}\ \bibnamefont {Guha}}, \bibinfo
  {author} {\bibfnamefont {Seth}\ \bibnamefont {Lloyd}}, \bibinfo {author}
  {\bibfnamefont {Lorenzo}\ \bibnamefont {Maccone}}, \bibinfo {author}
  {\bibfnamefont {Stefano}\ \bibnamefont {Pirandola}}, \ and\ \bibinfo {author}
  {\bibfnamefont {Jeffrey~H}\ \bibnamefont {Shapiro}},\ }\bibfield  {title}
  {\enquote {\bibinfo {title} {Quantum illumination with gaussian states},}\
  }\href@noop {} {\bibfield  {journal} {\bibinfo  {journal} {Physical review
  letters}\ }\textbf {\bibinfo {volume} {101}},\ \bibinfo {pages} {253601}
  (\bibinfo {year} {2008})}\BibitemShut {NoStop}%
\bibitem [{\citenamefont {De~Palma}\ and\ \citenamefont
  {Borregaard}(2018)}]{depalma:18}%
  \BibitemOpen
  \bibfield  {author} {\bibinfo {author} {\bibfnamefont {Giacomo}\ \bibnamefont
  {De~Palma}}\ and\ \bibinfo {author} {\bibfnamefont {Johannes}\ \bibnamefont
  {Borregaard}},\ }\bibfield  {title} {\enquote {\bibinfo {title} {Minimum
  error probability of quantum illumination},}\ }\href@noop {} {\bibfield
  {journal} {\bibinfo  {journal} {Physical Review A}\ }\textbf {\bibinfo
  {volume} {98}},\ \bibinfo {pages} {012101} (\bibinfo {year}
  {2018})}\BibitemShut {NoStop}%
\bibitem [{\citenamefont {Nair}\ and\ \citenamefont {Gu}(2020)}]{nair:20}%
  \BibitemOpen
  \bibfield  {author} {\bibinfo {author} {\bibfnamefont {Ranjith}\ \bibnamefont
  {Nair}}\ and\ \bibinfo {author} {\bibfnamefont {Mile}\ \bibnamefont {Gu}},\
  }\bibfield  {title} {\enquote {\bibinfo {title} {Fundamental limits of
  quantum illumination},}\ }\href@noop {} {\bibfield  {journal} {\bibinfo
  {journal} {arXiv preprint arXiv:2002.12252}\ } (\bibinfo {year}
  {2020})}\BibitemShut {NoStop}%
\bibitem [{\citenamefont {Braunstein}\ and\ \citenamefont
  {Caves}(1994)}]{braunstein:94}%
  \BibitemOpen
  \bibfield  {author} {\bibinfo {author} {\bibfnamefont {M.}~\bibnamefont
  {Braunstein}}\ and\ \bibinfo {author} {\bibfnamefont {C.M.}\ \bibnamefont
  {Caves}},\ }\href@noop {} {\bibfield  {journal} {\bibinfo  {journal} {Phys.
  Rev. Lett.}\ }\textbf {\bibinfo {volume} {72}} (\bibinfo {year}
  {1994})}\BibitemShut {NoStop}%
\bibitem [{\citenamefont {Uhlmann}(1992)}]{uhlmann:92}%
  \BibitemOpen
  \bibfield  {author} {\bibinfo {author} {\bibfnamefont {Armin}\ \bibnamefont
  {Uhlmann}},\ }\bibfield  {title} {\enquote {\bibinfo {title} {The metric of
  bures and the geometric phase},}\ }in\ \href@noop {} {\emph {\bibinfo
  {booktitle} {Groups and related Topics}}}\ (\bibinfo  {publisher}
  {Springer},\ \bibinfo {year} {1992})\ pp.\ \bibinfo {pages}
  {267--274}\BibitemShut {NoStop}%
\bibitem [{\citenamefont {Uhlmann}(1976)}]{uhlmann:76}%
  \BibitemOpen
  \bibfield  {author} {\bibinfo {author} {\bibfnamefont {Armin}\ \bibnamefont
  {Uhlmann}},\ }\bibfield  {title} {\enquote {\bibinfo {title} {The transition
  probability in the state space of a*-algebra},}\ }\href@noop {} {\bibfield
  {journal} {\bibinfo  {journal} {Reports on Mathematical Physics}\ }\textbf
  {\bibinfo {volume} {9}},\ \bibinfo {pages} {273--279} (\bibinfo {year}
  {1976})}\BibitemShut {NoStop}%
\end{thebibliography}%

\section*{Appendix 1}
The action of $\hat{U}_S(\theta)$ on $\Ket{\uparrow}$ is given by

\begin{align}
\hat{U}_S(\theta)\Ket{\uparrow}&=\frac{1}{\sqrt{2}}(e^{i\theta/2}\Ket{\phi^+_i}+ e^{-i\theta/2}\Ket{\phi^-_j}) \nonumber \\ 
&=\cos\frac{\theta}{2}\Ket{\uparrow}+ i\sin\frac{\theta}{2}\Ket{\downarrow} \approx \Ket{\uparrow}+ \frac{i\theta}{2}\Ket{\downarrow}.
\label{U}
\end{align}

At $\theta=0$, the operator $\hat{\rho}^1$ can be written
\begin{equation}
\hat{\rho}^1=\beta_+\Ket{\uparrow}\Bra{\uparrow}+\beta_0(\mathbb{1}-\Ket{\uparrow}\Bra{\uparrow}),
\end{equation} 
where
\begin{equation}
\beta_+=\frac{1-\eta}{d}+\eta,~~~\beta_0=\frac{1-\eta}{d}.
\end{equation}
Differentiating $\hat{\rho}^1$ with respect to $\theta$ and referring to Eq.(\ref{U}), we see
\begin{align}
\hat{\rho}'&=\eta \left( \frac{\partial (\hat{U}_S(\theta)\Ket{\uparrow})}{\partial \theta}\Bra{\uparrow}+\Ket{\uparrow}\frac{\partial (\Bra{\uparrow}\hat{U}_S(\theta))}{\partial \theta} \right) \nonumber \\
&=\frac{-i\eta}{2}\left(\Ket{\uparrow}\Bra{\downarrow}-\Ket{\downarrow}\Bra{\uparrow}\right)
\end{align}
Hence $\Braket{\uparrow|\hat{\rho}'|\uparrow}=\Braket{\downarrow|\hat{\rho}'|\downarrow}=0$, and so
\begin{equation}
R^{-1}_{\hat{\rho}^1}(\hat{\rho}')=\frac{\hat{\rho}'}{\beta_++\beta_0}=\frac{-i\eta d\left(\Ket{\uparrow}\Bra{\downarrow}-\Ket{\downarrow}\Bra{\uparrow}\right)}{4(1-\eta)+2\eta d}.
\end{equation}
And so we arrive at
\begin{equation}
\hat{\rho}'R^{-1}_{\hat{\rho}^1}(\hat{\rho}')=\frac{\eta^2 d\left(\Ket{\uparrow}\Bra{\uparrow}+\Ket{\downarrow}\Bra{\downarrow}\right)}{8(1-\eta)+4\eta d}.
\end{equation}
By taking the trace of this quantity we recover Eq. (\ref{peroutputphoton}).

\section*{Appendix 2}
The action of $\hat{U}_{S\otimes A}$ on $\Ket{k,m}$ is as follows:
\begin{align}
\hat{U}(\theta)_{S\otimes A}\Ket{k,m}=&\frac{1}{\sqrt{d}}\sum^{d}_{j=1} e^{2i\pi jk /d}U(\theta)\Ket{\phi_j}_S\Ket{\chi_{j+m}}_A  \\
=&\frac{1}{\sqrt{d}}\sum^{d}_{j\text{ odd}} e^{-2i\pi jk /d+i\theta/2}\Ket{\phi_j}_S\Ket{\chi_{j+m}}_A \\
&+\frac{1}{\sqrt{d}}\sum^{d}_{j\text{ even}} e^{-2i\pi jk /d-i\theta/2}\Ket{\phi_j}_S\Ket{\chi_{j+m}}_A  \nonumber
\end{align}

Noting that
\begin{equation}
\left(\frac{1+e^{i\pi j}}{2}\right)=\begin{cases}
1,~j\text{ even}\\
0,~j\text{ odd}
\end{cases} \text{ and } \left(\frac{1-e^{i\pi j}}{2}\right)=\begin{cases}
0,~j\text{ even}\\
1,~j\text{ odd}
\end{cases},
\end{equation}
we can write this as

%\begin{widetext}
%$$\mbox{U(\theta)\Ket{k,m}&= \frac{1}{\sqrt{d}}\sum^{d}_{j=1} e^{2i\pi jk/d}\left[e^{-i\theta/2}\left(\frac{1+e^{i\pi j}}{2}\right)+e^{i\theta/2}\left(\frac{1-e^{i\pi j}}{2}\right)\right]\Ket{\phi_j}_S\Ket{\phi_{j+m}}_A.\nonumber}$$
%\end{widetext}

%=&\frac{1}{\sqrt{d}}\sum^{d}_{j=1} \left[\left(\frac{e^{i\theta/2}+e^{-i\theta/2}}{2}\right)e^{2i\pi jk /d} +\left(\frac{e^{i\theta/2}-e^{-i\theta/2}}{2}\right)e^{2i\pi j(k+d/2) /d}\right]\Ket{\phi_j}_S\Ket{\phi_{j+m}}_A  \\ \nonumber
%=&\left(\mbox{cos}\frac{\theta}{2}\Ket{k,m}+i\mbox{sin}\frac{\theta}{2}\Ket{k+d/2,m}\right) \\ \approx &\left(\Ket{k,m}+\frac{i\theta}{2}\Ket{k+d/2,m}\right).

%\begin{widetext}
%\begin{align}
%U(\theta)\Ket{k,m}&= \frac{1}{\sqrt{d}}\sum^{d}_{j=1} e^{2i\pi jk/d}\left[e^{-i\theta/2}\left(\frac{1+e^{i\pi j}}{2}\right)+e^{i\theta/2}\left(\frac{1-e^{i\pi j}}{2}\right)\right]\Ket{\phi_j}_S\Ket{\chi_{j+m}}_A.\nonumber\\ \nonumber
%=&\frac{1}{\sqrt{d}}\sum^{d}_{j=1} \left[\left(\frac{e^{i\theta/2}+e^{-i\theta/2}}{2}\right)e^{2i\pi jk /d} +\left(\frac{e^{i\theta/2}-e^{-i\theta/2}}{2}\right)e^{2i\pi j(k+d/2) /d}\right]\Ket{\phi_j}_S\Ket{\chi_{j+m}}_A  \\ \nonumber
%=&\mbox{ cos}\frac{\theta}{2}\Ket{k,m}+i\mbox{ sin}\frac{\theta}{2}\Ket{k+d/2,m}.
%\end{align}
%\end{widetext}

\begin{align}
&\hat{U}(\theta)_{S\otimes A}\Ket{k,m}=\\
\frac{1}{\sqrt{d}}&\sum^{d}_{j=1} e^\frac{2i\pi jk}{d}\left[e^\frac{-i\theta}{2}\left(\frac{1+e^{i\pi j}}{2}\right)+e^\frac{i\theta}{2}\left(\frac{1-e^{i\pi j}}{2}\right)\right]\Ket{\phi_j}_S\Ket{\chi_{j+m}}_A\nonumber\\ \nonumber
=&\frac{1}{\sqrt{d}}\sum^{d}_{j=1} \left[\cos\frac{\theta}{2}e^{2i\pi jk /d} +\sin\frac{\theta}{2}e^{2i\pi j(k+d/2) /d}\right]\Ket{\phi_j}_S\Ket{\chi_{j+m}}_A  \\ 
=&\mbox{ cos}\frac{\theta}{2}\Ket{k,m}+i\mbox{ sin}\frac{\theta}{2}\Ket{k+d/2,m}.
\end{align}

This is the same form as Eq.\ref{U}. We choose to illuminate with any of the states $\Ket{k,m}$, say $\Ket{1,1}$. The derivation of the quantum Fisher information then takes the same form as in Case (1), with the substitutions $ d\rightarrow d^2,~\Ket{\uparrow} \rightarrow \Ket{1,1}, \Ket{\downarrow} \rightarrow \Ket{1+d/2,1}$. Hence we obtain Eq. (\ref{pairfisher}).

\section*{Appendix 3}

In Case 3, the illumination consists of a coherent-state probe over one mode in each of the + and - bases,
\begin{align}
\Ket{\psi_3}&=\Ket{\alpha}^+\Ket{\alpha}^-\\
&=\Ket{\sqrt{2}\alpha}^\uparrow\Ket{0}^\downarrow,
\end{align}
where $\uparrow$ and $\downarrow$ indicate modes defined by $\hat{a}_\uparrow=\frac{\hat{a}_++\hat{a}_-}{\sqrt{2}}$ and $\hat{a}_\downarrow=\frac{\hat{a}_+-\hat{a}_-}{\sqrt{2}}$. In this basis, $\theta$ parameterises a beam splitter coupling between the two modes, such that $\hat{U}_S(\theta)=e^{i\theta\hat{S}}$ where $\hat{S}\equiv\hat{a}_\uparrow\hat{a}^\dagger_\downarrow+\hat{a}^\dagger_\uparrow\hat{a}_\downarrow$. For comparison with Cases (1) and (2), we here take $\alpha=1/\sqrt{2}$.

The output state $\hat{\rho}^3(0)$, which features no coupling between these modes, may then be written as a product of a phase-averaged displaced thermal state in mode $\uparrow$ and a thermal state in mode $\downarrow$:
\begin{align}
\hat{\rho}^3(0)=\frac{1}{2\pi} \left(\int_{-\pi}^{\pi}\hat{G}_{\sqrt{2}\delta,\sigma}^\uparrow(\xi)\text{d}\xi\right)\otimes\hat{G}_{0,\sigma}^\downarrow.
\end{align}
Both of these states are phase-insensitive and are therefore diagonal in the Fock basis. To underscore this, we write $\hat{\rho}^3(0)\equiv\hat{\Lambda}$.
Next we write $\hat{\rho}^3(\varepsilon)$ in terms of this matrix and expand to second order in $\varepsilon$:
\begin{align}
\hat{\rho}^3(\varepsilon)&=e^{i\varepsilon\hat{S}}\hat{\Lambda}e^{-i\varepsilon\hat{S}}\\
&\approx(1+i\varepsilon\hat{S}-\frac{\varepsilon^2}{2}\hat{S}^2)\hat{\Lambda}(1-i\varepsilon\hat{S}-\frac{\varepsilon^2}{2}\hat{S}^2)\\
&\approx \hat{\Lambda}+i\varepsilon(\hat{S}\hat{\Lambda}-\hat{\Lambda}\hat{S})+\frac{\varepsilon^2}{2}(2\hat{S}\hat{\Lambda}\hat{S}-\hat{S}^2\hat{\Lambda}-\hat{\Lambda}\hat{S}^2).
\end{align}

Hence we write the quantity under the square root in Eq. (\ref{iqlimit})
\begin{equation}
\sqrt{\hat{\rho}(0)}\hat{\rho}(\varepsilon)\sqrt{\hat{\rho}(0)}\approx\hat{\Lambda}^2+\varepsilon\hat{P}+\varepsilon^2\hat{Q},
\end{equation}
where 
\begin{equation}
\hat{P}\equiv i\sqrt{\hat{\Lambda}}(\hat{S}\hat{\Lambda}-\hat{\Lambda}\hat{S})\sqrt{\hat{\Lambda}}
\end{equation}
and
\begin{equation}
\hat{Q}\equiv \frac{1}{2}\sqrt{\hat{\Lambda}}(2\hat{S}\hat{\Lambda}\hat{S}-\hat{S}^2\hat{\Lambda}-\hat{\Lambda}\hat{S}^2)\sqrt{\hat{\Lambda}}.
\end{equation}

To find the square root of this, we write
\begin{align}
\sqrt{\sqrt{\hat{\rho}(0)}\hat{\rho}(\varepsilon)\sqrt{\hat{\rho}(0)}}=\hat{\Lambda}+\varepsilon\hat{A}+\varepsilon^2\hat{B}
\end{align}

where
\begin{equation}
(\hat{\Lambda}+\varepsilon\hat{A}+\varepsilon^2\hat{B})^2=\hat{\Lambda}^2+\varepsilon\hat{P}+\varepsilon^2\hat{Q}.
\end{equation}
Collecting terms in $\varepsilon$ and $\varepsilon^2$, we find (writing in the Fock basis)
\begin{equation}
A_{ij}=\frac{P_{ij}}{\lambda^{(i)}+\lambda^{(j)}},
\end{equation}
and 
\begin{equation}
B_{ij}=\frac{Q_{ij}-\sum_k A_{ik}A_{kj}}{\lambda^{(i)}+\lambda^{(j)}},
\end{equation}
where $\lambda^{(i)}$ are the diagonal elements of $\hat{\Lambda}$.
Hence
\begin{equation}
\text{Tr}\left[\sqrt{\sqrt{\hat{\rho}(0)}\hat{\rho}(\varepsilon)\sqrt{\hat{\rho}(0)}}\right]=\text{Tr}\left[\hat{\Lambda}\right]+\varepsilon\text{Tr}\left[\hat{A}\right]+\varepsilon^2\text{Tr}\left[\hat{B}\right].
\end{equation}
Since $\hat{\Lambda}$ is a density matrix, $\text{Tr}\left[\hat{\Lambda}\right]=1$, and since the diagonal elements of $\hat{A}$ are all zero, $\text{Tr}\left[\hat{A}\right]=0$, as expected. Substituting into Eq. \ref{iqlimit}, we therefore find 

\begin{equation} \boxed{
I_Q^3=-8\text{Tr}\left[\hat{B}\right]}.
\label{I3}
\end{equation}

\end{document}